\documentclass[pra,showpacs,aps,twocolumn]{revtex4}
\usepackage{epsfig}
\begin{document}
\title{Squeezing concentration for Gaussian states with unknown
parameter}
\author{Radim Filip,
\footnote{email:filip@optics.upol.cz, tel:+420-68-5631572,
fax:+420-68-5224246}
and Ladislav Mi\v sta Jr.}
\affiliation{Department of Optics, Research Center for Optics,\\
Palack\' y University,\\
17. listopadu 50,  772~07 Olomouc, \\ Czech Republic}
\date{\today}

\begin{abstract}
A continuous-variable analog of the Deutsch's
distillation protocol locally operating with two
copies of the same Gaussian state is suggested.
Irrespectively of the impossibility of Gaussian state distillation, we reveal
that this protocol is able to perform a squeezing
concentration of the Gaussian states with {\em unknown} displacement.
Since this operation cannot be implemented using only single copy of the
state, it is a new application of the
distillation protocol which utilizes
two copies of the same Gaussian state.
\end{abstract}


\maketitle
\section{Introduction}

The exploitation of entangled states of quantum systems makes new
quantum information protocols possible such as quantum teleportation
\cite{Bennett93}, entanglement swapping \cite{Pan98} and quantum
cryptography \cite{Ekert91}. The common feature of
all these protocols is that they require to transmit the
entangled state from the
common source to the distant partners. In
practice, the transmission is always accompanied by losses that
contaminate the shared state and thus reduce the degree of entanglement.
Since the efficiency of the protocols is considerably dependent on the
degree of entanglement
shared by the partners, the question that naturally arises is whether
the partners can eliminate the influence of losses and to enhance the
entanglement employing only local
operations and classical communication (LOCC).
Such the protocols, commonly called distillation protocols, leading to the
concentration of entanglement contained in several copies of partially
entangled two-qubit states into a smaller number of singlet states has been
proposed \cite{Bennett96a,Bennett96b} and also experimentally realized
\cite{Pan01}.
Further, efficient distillation procedure converting any two-qubit
state with fidelity relative to maximally entangled state $F>1/2$ has been
suggested in \cite{Deutsch96}. Also entanglement swapping based
distillation of partially entangled pure states of two qubits has been
proposed in the literature \cite{Bose99}. The higher-dimensional
generalizations of the qubit distillation protocols have been given in
\cite{Horodecki99a} and \cite{Alber01}.

With increase of the interest in
quantum information processing with continuous variables
\cite{Braunstein}, \cite{Furusawa01} the natural need for distillation of
continuous-variable (CV) entangled states has arisen.
The existing proposals of CV distillation schemes allow either
to distill the bipartite Gaussian states at the expense of non-Gaussian
operations such as photon number measurement \cite{Duan00} or they are
capable to distill the bipartite non-Gaussian states \cite{Parker00}.
Unfortunately, any of these protocols is not satisfactory with respect
to the current experiment, in which only Gaussian operations with
Gaussian states are well managed. Recent results which apply to the
"purely" Gaussian distillation protocols have led to the conclusion
that no such a protocol exists. Namely, it was shown by means of the
Gaussian completely positive (CP) maps technique that any
trace-decreasing LOCC Gaussian CP map acting on known Gaussian state can be
replaced by trace-preserving LOCC Gaussian map CP \cite{Fiurasek02}. From that
it follows particularly, that a single copy of two-mode bipartite entangled
Gaussian state cannot be distilled to a more entangled state using only
LOCC Gaussian operations. Further it was proved, that the bipartite
symmetric two-mode Gaussian state cannot be distilled to more entangled state
with the aid of another identical copy of the state using only local
Gaussian operations \cite{Eisert02}. The general proof that Gaussian
states cannot be distilled by local Gaussian operations and classical
communication was given in \cite{Giedke02}. Having these facts
in mind the question that naturally arises is whether there exists some
Gaussian operation on bipartite entangled Gaussian state that two distant
observers employing only local Gaussian operations wish to carry out that
can be performed only at the expense of utilization of more than single
copy of the state. This is the subject of the present article.
The paper is organized as follows. In Sec. II we investigate the
continuous-variable analog of the Deutsch's qubit distillation protocol.
The example of Gaussian operation requiring 'collective' local operations
on two copies of entangled Gaussian state is given in Sec. III. Finally,
Sec. IV contains conclusion.
\section{CV analog of Deutsch's protocol}

\begin{figure}
\centerline{\psfig{width=6.0cm,angle=0,file=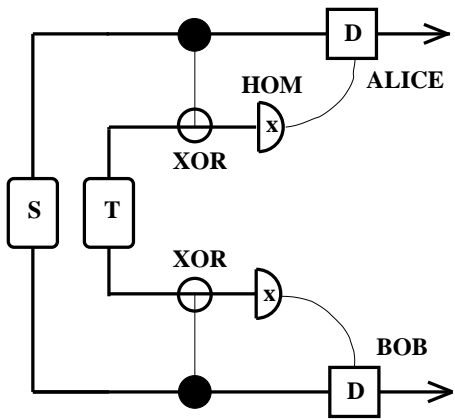}}
\caption{Two-mode squeezing concentration scheme: S -- source pair of the modes, T -- target pair of the modes, HOM -- homodyne detection, D -- 
coordinate unitary displacement.}
\end{figure}

The properties of the CV analog of the Deutsch's protocol \cite{Deutsch96},
which will be described below, are best studied on the example of the
displaced two-mode squeezed vacuum state
\begin{eqnarray}\label{nopa}
|x_{0},\sigma_{+},\sigma_{-}\rangle=\frac{1}{\sqrt{\pi\sqrt{\sigma_{+}\sigma_{-}}}}
\int\!\!\!\int_{-\infty}^{\infty}dxdy\times\nonumber\\
\exp\left[-\frac{(x-y-x_{0})^{2}}{4\sigma_{-}}
-\frac{(x+y-x_{0})^{2}}{4\sigma_{+}}
\right]|x\rangle_{A}|y\rangle_{B},
\end{eqnarray}
where $\sigma_{+}=\langle[\Delta(X_{A}+X_{B})]^{2}\rangle=
\mbox{e}^{2r_{1}}$, $\sigma_{-}=\langle[\Delta(X_{A}-X_{B})]^{2}\rangle=
\mbox{e}^{-2r_{2}}$ $r_1$, $r_2$ are squeezing parameters;
$x_{i}$ and $p_{j}$, $i,j=A,B$, $[X_{i},P_{j}]=i\delta_{ij}$ are standard
position and momentum quadrature operators of modes $A$ and $B$, respectively;
$x_{0}$ is an {\em unknown} coherent displacement of quadrature $X_{A}$.
Note, that as $\sigma_{-}$ decreases, $\sigma_{+}$ increases, and this state
approaches the well-known Einstein-Podolsky-Rosen state. The state
(\ref{nopa}) can be prepared by mixing of momentum-squeezed vacuum state
with squeezing $r_1$ and position-squeezed vacuum state with squeezing
$r_2$ at a balanced beamsplitter \cite{Furusawa01}, followed by unitary
displacement operation $U=\mbox{exp}(-ix_{0}P_{A})$ on mode $A$.

Let us assume that two identical copies, conventionally called source
(S) and target (T), of the state (\ref{nopa}) where $x_{0}=0$ are distributed
between two observers, Alice (A) and Bob (B), as is depicted in Fig.~1.
The natural extension of the Deutsch's qubit distillation protocol
\cite{Deutsch96} to CV then consists of (i) local QND measurement of
source position quadrature (CV analog of XOR-gate)
\begin{equation}
|x\rangle_{S}|x'\rangle_{T} \rightarrow
|x\rangle_{S}|x'-x\rangle_{T},
\end{equation}
performed both by Alice and by Bob, followed by (ii) local homodyne
measurement of the target quadratures $X_{{T}_{A}}$ and $X_{{T}_{B}}$.
Alice and Bob then can communicate their measurement outcomes via
classical channel and, as in the original protocol, discard both the pairs
if the outcomes do not coincide. A straightforward calculation reveals,
however, that irrespectively of the measurement outcomes all the output
states of the protocol can be brought into the single state that is
independent on the measurement outcomes, by applying two local unitary
displacement transformations
\begin{eqnarray}\label{displ}
x\rightarrow x+{\cal X}/2,\hspace{0.3cm} y\rightarrow y+{\cal Y}/2
\end{eqnarray}
on Alice's and Bob's side, if the measurement outcomes obtained by them
are ${\cal X}$ and ${\cal Y}$, respectively. Thus the selection of the
subensemble of the output states based on the results of target measurements
is not useful and the protocol becomes {\em deterministic}. This property of
the CV analog of the Deutsch's scheme is in sharp contrast with the original
protocol for qubits. It has been shown recently \cite{Fiurasek02}, that this
behaviour is characteristic to all Gaussian distillation protocols for apriori
{\em known} input Gaussian states, i.e. that any such 'purely' Gaussian
distillation protocol would be deterministic.

After application of this CV protocol and displacements (\ref{displ}), one
finds the output state of the form (\ref{nopa}) with the variances changed
according to the rule $\tilde{\sigma}_{\pm}=\frac{\sigma_{\pm}}{2}$.
Irrespectively of the increase of the position correlations, the marginal
purity $P=2\sqrt{\sigma_{+}\sigma_{-}}/(\sigma_{+}+\sigma_{-})$
of outgoing pure state remains unchanged and the entanglement in the state
(\ref{nopa}) is not enhanced. In order to decrease the marginal purity $P$
and consequently to obtain more entangled state, we would need different
behaviour in these variances, for example, $\sigma_{-}$ decreases and
$\sigma_{+}$ remains constant at a time.

Surprisingly, our CV procedure can be substituted by the {\em unconditional}
scheme employing only {\em single copy} of the source state. It
consists of two single-mode squeezers, performing the following squeezing
transformations ${\tilde X_{A}}=\frac{1}{\sqrt{2}}X_{A}$ and
${\tilde X_{B}}=\frac{1} {\sqrt{2}} X_{B}$ on Alice's and Bob's side.
This procedure is capable of transformation of nonsymmetrically entangled
states with $\sigma_{+}\sigma_{-}>1$ to symmetrical ones, having
$\sigma_{+}\sigma_{-}=1$, and vice versa, as has been previously
discussed in \cite{Bowen01}. Thus, due to the equivalence between
aforementioned two-copy and single-copy schemes, one can have doubt about
the usefulness of our CV procedure based on operations on two copies of
the input state. The question that naturally arises in this context is, whether the
utilization of two copies of a two-mode Gaussian state enables us to perform
operation, which cannot be equivalently carried out only on single copy of the
state. In the following Section, we demonstrate that our CV analog of the
Deutsch's protocol can be useful in this respect, particularly for
squeezing
concentration of the Gaussian states (\ref{nopa}) with an {\em unknown}
displacement $x_{0}$.

\section{Squeezing concentration}
\begin{figure}
\vspace{0cm}
\centerline{\psfig{width=7.0cm,angle=0,file=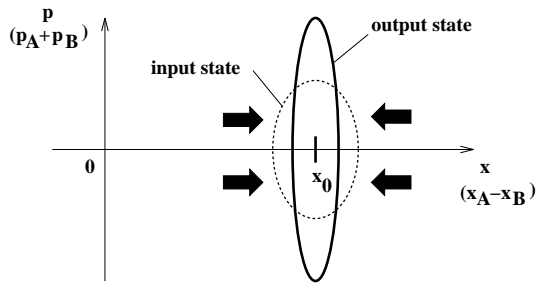}}
\vspace{0cm}
\caption{Squeezing concentration procedure: the noise ellipse is squeezed 
without a change of {\em unknown} mean value $x_{0}$.}
\end{figure}

We start from a simple example illustrating the usefulness of the
protocol discussed in the previous Section in manipulating with an
{\em unknown} Gaussian state. First, we analyse a local part of the
protocol, for example, on the Alice's side.
Let us consider the following single-mode
pure Gaussian state
\begin{equation}\label{single}
|x_{0},\sigma_{X}\rangle=(2\pi\sigma)^{-1/4}\int_{-\infty}^{\infty}\exp
\left[-\frac{(x-x_{0})^{2}}{4\sigma_{X}}\right]|x\rangle dx
\end{equation}
where $x_{0}=\langle X\rangle$ is an {\em unknown} parameter representing
a coherent signal and $\sigma_{X}=\langle (\Delta X)^{2}\rangle=
\mbox{e}^{-2r}/2$, $r$ is the squeezing parameter. This state can be
prepared by displacing the squeezed vacuum state by the value $x_{0}$.
The displacement is often used to encode information
into the quantum state, for example, in the CV dense coding
\cite{Braunstein00}.
Since the transmission channel introduces extra noise into the
state (\ref{single}), it is desirable to reduce the noise affecting the
transmitted information before subsequent processing.
It can be achieved by means of appropriate
squeezing in the position quadrature. Employing only single copy of the
state (\ref{single}), we can simply squeeze the position fluctuations in
the single-mode squeezer, however, in this case also the
value of {\em unknown} parameter $x_{0}$ is changed. The reason is that
single-mode squeezer reduces both the mean value $\langle X\rangle$ and
the variance $\sigma=\langle (\Delta X)^{2}\rangle$ of the position
quadrature. If we do not know the displacement $x_{0}$, then we cannot
restore the mean value to the original one without errors. Because the
information is encoded in the mean value $\langle X\rangle$, its preservation
during squeezing is important. On the other hand, knowing the parameter $x_{0}$,
we can restore the initial mean value by a suitable displacement.

However, if we consider two copies of the state (\ref{single}),
we are able to squeeze the position fluctuations without changing
{\em unknown} mean value  $\langle X\rangle=x_{0}$, as is depicted in
Fig.~2. It can be achieved by above discussed protocol, employing, for
instance, only Alice's part of the scheme outlined in Fig.~1. The
procedure produces single copy of the state (\ref{single}) with
new reduced variance $\tilde{\sigma}=\frac{\sigma}{2}$ and without
changing the mean value $\langle X\rangle$. Since it cannot be achieved
employing only single copy of the state (\ref{single}), this example
illustrates a new application of the CV analog of the distillation protocol
in manipulation with {\em unknown} Gaussian states.

In general, this protocol can be looked at as a state transformation of
the source Wigner function $W_{S}(x,p)$ with the help of target Wigner
function $W_{T}(x,p)$. After QND measurement, position measurement on the
target mode and subsequent displacement, we arrive at the output
source Wigner function
\begin{eqnarray}\label{wigner1}
\tilde{W}_{S}(x,p)&=&\int\!\!\!\int_{-\infty}^{\infty}W_{S}(x-{\cal
X}/2,p')\times\nonumber\\
& & W_{T}(x+{\cal X}/2,p-p')dp'd{\cal X}.
\end{eqnarray}
If we are interested in the position and momentum distributions of the
outgoing source mode separately, we can integrate the Wigner function
(\ref{wigner1}) over momentum $p$ and position $x$, respectively
and obtain the following marginal distributions
\begin{eqnarray}\label{marg}
\tilde{p}_{S}(x)&=&\int_{-\infty}^{\infty}p_{S}(x-{\cal
X}/2)p_{T}(x+{\cal X}/2)d{\cal X},\nonumber\\
\tilde{p}_{S}(p)&=&\int_{-\infty}^{\infty}p_{S}(p')p_{T}(p-p')dp',
\end{eqnarray}
where $p_{S}$ and $p_{T}$ are input source and target marginal
distributions, respectively. Assuming both the source and target state
to be in the same Gaussian state with marginal distributions
\begin{eqnarray}
p_{S}(x)=p_{T}(x)&=&\frac{1}{\sqrt{2\pi\langle(\Delta X_{S})^{2}\rangle}}
\exp\left[-\frac{(x-x_{0})^{2}}{2\langle(\Delta
X_{S})^{2}\rangle}\right],\nonumber\\
p_{S}(p)=p_{T}(p)&=&\frac{1}{\sqrt{2\pi\langle(\Delta P_{S})^{2}\rangle}}
\exp\left[-\frac{p^{2}}{2\langle(\Delta
P_{S})^{2}\rangle}\right],\nonumber\\
\end{eqnarray}
substituting them into the formulas (\ref{marg}) and performing the
integrations, we obtain that the {\em unknown} mean value
$\langle{\tilde X}_{S}\rangle=\langle X_{S}\rangle=x_{0}$
is preserved, whereas the position fluctuations are squeezed to half of
initial value
\begin{equation}
\langle(\Delta\tilde{X}_{S})^{2}\rangle=
\frac{\langle(\Delta X_{S})^{2}\rangle}{2},\hspace{0.3cm}
\langle(\Delta\tilde{P}_{S})^{2}\rangle=
2\langle(\Delta P_{S})^{2}\rangle
\end{equation}
around this mean value. Consequently, due to principle of complementarity,
the momentum fluctuations are enhanced. Employing $N$ identical copies of
the same Gaussian state, we are able to squeeze the position
fluctuations to $1/N$ of initial value, without changing the mean value of
the position quadrature. Naturally, similar procedure can be constructed
for the squeezing of momentum fluctuations. Note, that a discrete-variable
analogue of this procedure has been discussed as the qubit purification
protocol \cite{Bowmeester00}.

Let us apply now the idea of this procedure to the two-mode case,
considering two copies of the state (\ref{nopa}), for simplicity.
If the parameter $x_{0}$ is known, then we can utilize the Bowen's
``concentrating'' procedure and manipulate with two-mode correlations
on single copy \cite{Bowen01}. If, however, the value $x_{0}$ is a priori
not known, then we are not able to coherently manipulate with the
variance $\langle[\Delta(X_{A}-X_{B})]^{2}\rangle$, without changing
the unknown parameter $x_{0}$. On the other hand, employing entire
two-mode setup depicted in Fig.~1, we are able to squeeze coherently
the fluctuations ${\tilde\sigma}_{+}=\sigma_{+}/2$,
${\tilde\sigma}_{-}=\sigma_{-}/2$ again without changing the {\em unknown}
parameter $x_{0}$. In terms of Wigner functions, our concentrating
procedure can be expressed, in analogy with the single-mode case
(\ref{wigner1}), by the formula
\begin{widetext}
\begin{eqnarray}\label{wigner2}
\tilde{W}_{S}(x_{A},p_{A},x_{B},p_{B})&=&
\int\!\!\!\int\!\!\!\int\!\!\!
\int_{-\infty}^{\infty}
W_{S}(x_{A}-{\cal X}/2,p'_{A},x_{B}-{\cal Y}/2,p'_{B})\times\nonumber\\
& &W_{T}(x_{A}+{\cal X}/2,p_{A}-p'_{A},x_{B}+{\cal
Y}/2,p_{B}-p'_{B})dp'_{A}dp'_{B}d{\cal X}d{\cal Y}.
\end{eqnarray}
\end{widetext}
If we are interested separately in position correlations and momentum
correlations between Alice and Bob, we can integrate the function
(\ref{wigner2}) over momentum variables $p_{A}$, $p_{B}$ to obtain
joint position distribution
\begin{eqnarray}
\tilde{p}_{S}(x_{A},x_{B})=\int\!\!\!\int_{-\infty}^{\infty}p_{S}(x_{A}-{\cal
X}/2,x_{B}-{\cal Y}/2)\times\nonumber\\
p_{T}(x_{A}+{\cal X}/2,x_{B}+{\cal Y}/2)d{\cal X}d{\cal{Y}}
\end{eqnarray}
or we can integrate over position variables $x_{A}$, $x_{B}$ to obtain
joint momentum distribution
\begin{eqnarray}
\tilde{p}_{S}(p_{A},p_{B})
&=&\int\!\!\!\int_{-\infty}^{\infty}p_{S}(p'_{A},p'_{B})\times\nonumber\\
& &p_{T}(p_{A}-p'_{A},p_{B}-p'_{B})dp'_{A}dp'_{B}.
\end{eqnarray}
Let us assume, that the source and target have the same joint
probability distributions of the form
\begin{eqnarray}\label{twomode}
p({\bf x})&=&\frac{1}{2\pi\sqrt{\det {\bf V_{X}}}}
\exp\left[-({\bf x}-{\bf x}_{0})^{T}{(2{\bf V}_{X})}^{-1}({\bf x}-{\bf
x}_{0})\right],\nonumber\\
p({\bf p})&=&\frac{1}{2\pi\sqrt{\det {\bf V_{P}}}}
\exp\left[-{\bf p}^{T}{(2{\bf V}_{P})}^{-1}{\bf p}\right],
\end{eqnarray}
where ${\bf x}=(x_{A},x_{B})$, ${\bf x}_{0}=(x_{0},x_{0})$,
${\bf p}=(p_{A},p_{B})$, $T$ denotes
operation of transposition and where
\begin{equation}
({\bf V}_{X})_{ij}=\langle \Delta X_{i}\Delta X_{j}\rangle,
\hspace{0.3cm} ({\bf V}_{P})_{ij}=\langle \Delta P_{i}\Delta P_{j}\rangle
\end{equation}
are the elements of position and momentum variance matrices, where
$\Delta X_{i}=X_{i}-\langle X_{i}\rangle$, $\Delta P_{j}=P_{j}-\langle
P_{j}\rangle$, $i,j=A,B$. The simple calculation then yields the
output probability distributions of the same form as in
Eq.~(\ref{twomode}), however, with the transformed variance matrices
${\tilde{\bf V}}_{X}={\bf V}_{X}/2$ and ${\tilde{\bf V}}_{P}=2{\bf V}_{P}$.
Thus, the correlations in positions increase as follows
\begin{equation}
\langle[\Delta(\tilde{X}_{A}-\tilde{X}_{B})^{2}]\rangle=
\frac{\langle[\Delta({X}_{A}-{X}_{B})^{2}]\rangle}{2},
\end{equation}
whereas the mean values $\langle X_{A}\rangle$ and $\langle X_{B}\rangle$
are preserved. On the other hand, the momentum anticorrelations decrease
\begin{equation}
\langle[\Delta(\tilde{P}_{A}+\tilde{P}_{B})]^{2}\rangle=
2\langle[\Delta({P}_{A}+{P}_{B})]^{2}\rangle,
\end{equation}
and thus the entanglement and total entropy of the source state are preserved.
This example illustrates that although the CV analog of the Deutsch's
distillation protocol is not useful for enhancement of entanglement, it
can be useful when one wishes to locally concentrate the two-mode squeezing
in a partially {\em unknown} Gaussian state.

We have analysed the CV analog of specific distillation protocol for
two copies of pure two-mode  Gaussian state. We have demonstrated
that it is only able to manipulate with squeezing, while the
entanglement is preserved. Irrespectively to impossibility of
entanglement increasing, it can be useful from another point of view.
Employing two-copies of a two-mode Gaussian state displaced in the
position by an {\em unknown} value, we can utilize this local-operation
protocol to enhance the particular correlations, without changing the
{\em unknown} position displacement. Because this procedure cannot be
implemented if we have only single copy of this state, we have revealed
a new application of the CV analog of the discrete-variable
distillation protocol.

\medskip
\noindent {\bf Acknowledgments}
The authors would like to thank J. Fiur\' a\v{s}ek for fruitful discussions.
The work was supported by the project LN00A015 and CEZ:J14/98 of the
Ministry of Education of Czech Republic and
by the EU grant under QIPC project
IST-1999-13071 (QUICOV).


\end{document}